{\ }

\centerline{\bf A Simple Expression for the Cold Compression Curve}

\vskip1cm

\centerline{\bf V. Celebonovic}

\vskip.5cm

\centerline{\it Institute of Physics, Pregrevica 118, 11080 Zemun-Beograd,
Yugoslavia}

\centerline{\it E-mail celebonovic@exp.phy.bg.ac.yu}

\centerline{\it {\rm and:} E-mail vladanc@mrsys1.mr-net.co.yu}

\vskip1cm

\noindent Abstract: The aim of this contribution is to present expressions for
	the bulk modulus of a material and its pressure derivative obtained by using
the semiclassical theory of dense matter proposed by P. Savic and R. Kasanin.
Some possibilities for the application of these expressions are briefly
discussed.

\vskip.5cm

\centerline{\bf Introduction}

\vskip.5cm

The term "equation of state" (EOS) of a system denotes any kind of
relationship, established experimentally or theoretically, between the
parameters of state
of the system under consideration. In the particular case of a
thermomechanical system, the EOS has the general form $f(p, V, T) = 0$. The
knowledge of this function is of fundamental importance in various physical
and astrophysical problems (for recent examples see, for instance, Holzapfel,
1996; Schulte and Holzapfel, 1996; Schaab, Weber {\it et al.} 1996).

A simplified version of the EOS, which does not take into account the influence
of the temperature is called the cold compression curve (CCC). Strictly
speaking, the CCC is a $T= 0$K isotherm of dense matter, but, in practice, it
can be applied at low temperatures. In theoretical studies, the form of a
cold compression curve is a consequence of the assumptions of the underlying
theory ( Holzapfel, 1996 contains an instructive list). One of the simplest
forms of a CCC is the so-called Murnaghan EOS, which amounts to

$$P(\rho) = (B/B^{\prime})[(\rho_0 / \rho)^{-B^{\prime}} -1] \eqno{(1{\rm a})}$$

\noindent In this expression, $B$ and $B^{\prime}$ denote the isothermal bulk
modulus of a material under standard conditions, and its first pressure
derivative. These two functions are obviously material dependent, and their
knowledge is crucial for all applications of EOS. The bulk modulus is defined
as

$$B= -V(\partial P / \partial V)_T = \rho(\partial P / \partial \rho)_T
\eqno{(1)}$$

The aim of this contribution is to propose simple expressions for the
isothermal bulk modulus and its pressure derivative. All the calculations will
be performed within a particular semi-classical theory of the behaviour of
materials under high pressure, proposed by P. Savic and R. Kasanin in the
	early sixties (Savic and Kasanin, 1962/65). For recent reviews of their theory
see Savic and Celebonovic (1994); Celebonovic (1995).

\vskip.5cm

\centerline{\bf The Calculations}

\vskip.5cm

The necessary "ingredient" for the calculation of $B$ and $B^{\prime}$ is the
function $\partial P / \partial \rho$. In the theory proposed by Savic and
Kasanin this function has the following form.

$$\partial P / \partial \rho = (N_A e^2 / 9 A) Q \eqno{(2)}$$

\noindent where

$$Q = (4 / a_i) f_i (a_i) - f_i^{\prime} (a_i) \eqno{(2{\rm a})}$$

The symbol $f_i(a_i)$ denotes the function

$$f_i (a_i) = C_i + B_i \ {\rm exp} (\gamma_i z_i) \eqno{(3{\rm a})}$$

\noindent with

$$a_i = (A / 8 N_A \rho_i)^{1/3} \eqno{(3{\rm b})}$$

\noindent and

$$z_i = (1 - a_i^* / a_i) / (1 - \alpha_i^{-1/3}) \eqno{(3{\rm c})}$$

Details of derivations of these equations are avaliable in the original
publications of Savic and Kasanin. The meaning of various symbols is as
follows: $A$ and $N_A$  denote the atomic mass of the material and Avogadro's
number, $a_i^*, \ B_i, \ C_i, \ \gamma_i, \ \alpha_i$ are constants (whose
numerical values are known within the theory) in  the i-th phase  of a
material subdued to high pressure. Physically, $a_i$ is the mean
inter-particle distance in the i-th phase.

Inserting eq. (3a) into (2a), it follows that

$$Q_i = (4 / a_i) [C_i + B_i \ {\rm exp} (\gamma_i z_i)] - \gamma_i B_i \ {\rm
exp} (\gamma_i z_i) \partial z_i / \partial a_i \eqno{(4{\rm a})}$$

\noindent Expressing $\partial z / \partial a$ as $(\partial z / \partial
\rho) (\partial \rho / \partial a)$ and using eqs. (3b) and (3c), one gets

$$\eqalign{&Q_i = 2 (N_A \rho_i / A)^{1/3} [4(C_i + B_i \ {\rm exp} (\gamma_i
z_i) - B_i \gamma_i (\rho_i / \rho_i^*)^{1/3}\cr
&(1 - \alpha_i^{-1/3})^{-1} \ {\rm exp} (\gamma_i z_i)]\cr} \eqno{(5)}$$

\noindent Finally, the bulk modulus of the i-th phase of a material having the
atomic mass $A$ and density $\rho$ is given by

$$\eqalign{&B = 2 \rho_i (N_A e^2 / 9 A) (N_A \rho_i / A)^{1/3} [4 (C_i + B_i
{\rm exp} (\gamma_i z_i) -\cr
&B_i \gamma_i (1 - \alpha_i^{-1/3})^{-1}(\rho_i / \rho_i^*)^{1/3} \ {\rm exp}
(\gamma_i z_i)]\cr} \eqno{(6)}$$

\noindent The pressure derivative of $B$ can be calculated from eqs. (1) and
(2). It thus follows that

$$\eqalign{&B^{\prime} = \partial B / \partial P = (\partial / \partial P)
(\rho (\partial P / \partial \rho)) = \partial / \partial P [\rho N_A e^2 Q / 9
A] =\cr
&= 1 + \rho (N_A e^2 / 9 A) \partial Q / \partial P = 1 + (\rho / Q) \partial Q
/ \partial \rho\cr} \eqno{(7)}$$

\noindent and

$$B^{\prime} \cong 1 + 8 C_i N_A^{1/3} / (24 C_i N_A^{1/3} + \ll 2 \gg) + \ll
3 \gg + ... \eqno{(8)}$$

\noindent where $\ll 2 \gg$ and $\ll 3 \gg$ denote the number of terms omitted
due to 	space limitations.

\vskip.5cm

\centerline{\bf Conclusions}

\vskip.5cm

In this note we have briefly presented  a theoretical algorithm for the
determination of the isohermal bullk modulus and its first pressure derivative
within a particular semi-classical theory of dense matter. This is a distinct
advantage over the existing  literature in the high-pressure field, where $B$
and $B^{\prime}$ are usually derived by fitting the experimental p-V data to
some previously chosen form of the EOS.

The results obtained in this note give the possibility of  refining the models
of the internal structure of cold dense astrophysical objects (such as
planets satellites or asteroids)  as well as theoretically following the
changes of  compressibility of  laboratory specimen undergoing phase
transitions under high pressure. Details will be published elsewhere.

\vskip1cm

\noindent References

\vskip.5cm

\noindent Celebonovic, V.: 1995, {\it Bull. Astron. Belgrade},  {\bf 151},
     37,  avaliable also as preprint astro-ph/9603135.

\vskip3mm

\noindent Holzapfel,  W. B.: 1996,  {\it Rep. Progr. Phys.} {\bf 59},  29.

\vskip3mm

\noindent Savic,  P. and Celebonovic,  V.: 1994,  in {\it Amer. Inst. Phys. 
Conf. Proc.} {\bf 309},  53,  AIP Press,  New York. 

\vskip3mm

\noindent Savic,  P. and Kasanin,  R.: 1962/65,  The Behaviour of Materials 
Under High Pressure I-IV,  Ed. SANU,  Beograd. 

\vskip3mm

\noindent Schaab,  C.,  Weber,  F.,  Weigel,  M. K. and Glendenning,  N. K.: 
1996, preprint astro-ph/9603142. 

\vskip3mm

\noindent Schulte,  O. and Holzapfel,  W. B.: 1996,  {\it Phys. Rev. B} {\bf 
53}, 569. 

\bye